\documentclass[twocolumn,showpacs,prl,aps,superscriptaddress]{revtex4}
\usepackage{ulem}
\usepackage{array}
\usepackage{amsmath}
\usepackage{graphicx}
\usepackage{dsfont}
\begin{document}

\title{Generic Bell inequalities for multipartite arbitrary dimensional
  systems}

\author{W. Son}

\affiliation{School of Mathematics and Physics, Queen's University,
  Belfast BT7 1NN, United Kingdom}

\author{Jinhyoung Lee}
\affiliation{Department of Physics, Hanyang University, Seoul 133-791,
  Korea}

\affiliation{Quantum Photonic Science Research Center, Hanyang
  University, Seoul 133-791, Korea}

\author{M. S. Kim}
\affiliation{School of Mathematics and Physics, Queen's University,
  Belfast BT7 1NN, United Kingdom}

\date{\today}

\begin{abstract}
  We present generic Bell inequalities for multipartite arbitrary
  dimensional systems. The inequalities that any local realistic
  theories must obey are violated by quantum mechanics for even
  dimensional systems. A large set of variants are shown to naturally
  emerge from the generic Bell inequalities. We discuss particular
  variants of Bell inequalities, that are violated for all the systems
  including odd dimensional systems.
\end{abstract}


\maketitle

Quantum nonlocality is the most significant evidence of physical
observations that cannot be explained by theories based upon local
realism.  Local realism is rooted in the classical view of measurement,
namely that an observation on one of a pair of subsystems cannot affect
the other system faster than the speed of light. In fact, since the
advent of quantum mechanics, the nonclassical implications have given
rise to fundamental questions on the nature of the act of measurement.
In quantum mechanics, a measurement does not provide a preexisting value
of the system, rather, it is a manifestation of the state of the probed
system and the probing apparatus, as advocated by Mermin
\cite{Mermin93}.

As early as 1964, Bell \cite{Bell64} proved that local realism implies
constraints on a correlation of measurements between two separate
systems.  These constraints are incompatible with the quantitative
predictions by quantum theory in the case of two coupled spin-1/2
particles.  These constraints, expressed as so-called Bell inequalities,
are of paramount importance in the conceptual foundations of quantum
mechanics.  But these are idealised experiments with archetypal
non-classical system. A key question remains whether a complex system of
high-dimensional quantum subsystems could eventually simulate a
pseudo-classical system that does not contradict local realism.

Since this startling discovery \cite{Bell64}, investigating Bell theorem
for a general system has been regarded as one of the most important
challenges in quantum mechanics and quantum information science (QIS)
\cite{Werner,Garg80,Mermin90,Ardehali92,Belinski93,Gisin92,
  Kaszlikowski00,Collins02,Acin04,Greenberger89,Lee04,Cerf02,Cabello02}.
The motivation is obvious from a scientific and technological viewpoint.
Firstly, proving Bell theorem for a general system, would show that
quantum physics would apply to macroscopic complex systems.  Secondly,
for QIS to outperform the acquisition, manipulation and transmission of
information over its classical counterpart, the property of nonlocality
is closely related to its extraordinary power.  In fact, the
manipulation of a complex quantum system rather than a simple one has
practical advantages. Gaining access to the states is easier and more
efficient (for example, efficient cluster state QIS \cite{hans} and
increased security in high-dimensional quantum cryptography
\cite{Cerf02b}).  It follows that a nonlocality test for such a complex
system is highly desirable.  Thirdly, the controllability of quantum
operations depends on the nature of the system. Certain logical
operations are relatively easy for one system but impossible or
difficult for another.  It is then essential to understand the nonlocal
properties of different systems in order to couple them together, to
arrange interfaces between the systems.

Suppose that measurements are performed locally on $N$ subsystems.  On
each subsystem, one out of {\em two} observables is measured, bearing
$d$ outcomes each, in order to consider a Bell inequality in the
composite system of $(N, d)$ \cite{Werner}.  Garg and Mermin formulated
marginal probabilities predicted by quantum mechanics for $(2, d)$ and
investigated if they can be derived from higher-order joint
probabilities \cite{Garg80}.  This investigation suggested loss of
incompatibility between quantum mechanics and local realism in the limit
of $d\rightarrow\infty$.  Kaszliowski {\em et al.} \cite{Kaszlikowski00}
argued that loss of incompatibility was due to restrictions in the type
of measurement. They suggested the full use of the $d$-dimensional
Hilbert space to address this issue.  Later, Collins {\em et al.}
derived Bell inequalities for high $d$ values, which can exhibit the
incompatibility for $d\rightarrow\infty$ \cite{Collins02}.  These works
together with an information-theoretic approach of the Bell theorem
\cite{Braunstein88} are all for a composite system of $(2, d)$.  For a
multipartite spin-1/2 system of $(N, 2)$, Mermin \cite{Mermin90},
Ardehali \cite{Ardehali92}, and Belinskii and Klyshko \cite{Belinski93}
formulated Bell inequalities based on the statistical properties of the
Greenberger, Horne and Zeilinger (GHZ) nonlocality \cite{Greenberger89}.
In summary, we have seen proofs of the Bell theorem for systems of $(2,
d)$ and $(N, 2)$ for any $N$ and $d$. However, note that there are no
proofs for arbitrary $(N, d)$ despite its prime importance, as this will
complete the proof to rule out any local realistic view which has been
driven into the corner to stand for measurements having a continuum of
values for a continuum of particles.  Very recently, Lee {\em et al.}
showed Bell theorem without inequalities for multipartite
multidimensional systems in generalized GHZ states \cite{Lee04}. In this
Letter we generalize the Mermin-Ardehali Bell inequality for a system of
arbitrary $(N, d)$.  However, we neither suggest to find all the Bell
inequalities nor their nonlocality condition.  We do not consider many
measurement settings other than two, either. This is because our
interest is the currently important issue, a proof of the Bell theorem
for arbitrary $(N, d)$ system, which has been a long-awaited problem.

{\em Generic Bell inequalities.--} Before investigating general cases,
in order to understand the principal ideas, we consider a tripartite
arbitrary dimensional system which is already significant. Consider
three observers and allow each to independently choose one of two
variables. The variables are denoted by $A_j$ and $B_j$ for $j$-th
observer. Each variable takes, as its value, an element in the set
$S=\{1, \omega, \dots, \omega^{d-1}\}$ where the elements of $S$ are the
$d$-th roots of unity over the complex field. With these variables and
their powers, $A_j^n$ and $B_j^n$, we propose a generic Bell function,
$\mathcal{B}$,
\begin{equation}
\label{eq:bellfun}
\mathcal{B} =
\frac{1}{2^3}\sum_{n=1}^{d-1}\left\langle\prod_{j=1}^3\left(A_j^n +
\omega^{n/2}B_j^n\right)\right\rangle + c.c.
\end{equation}
where $c.c.$ stands for complex conjugate. The symbol $\langle \cdot
\rangle$ is introduced to denote the statistical average over many runs
of the experiment. It is remarkable that the higher-order correlation
functions appear in our Bell function. If $d=2$, our generic Bell
function is reduced to Mermin's \cite{Mermin90}. In the classical view
of the statistical average, the local realism implies that the values
for the variables are predetermined, before measurement, by local hidden
variable $\lambda$: $A_j(\lambda)$ and $B_j(\lambda)$. The correlation
among the variables is the statistical average over $\lambda$, i.e.,
\begin{eqnarray*}
  \label{eq:cfhvd}
  \int d\lambda ~\rho(\lambda) \prod_{j=1}^3 V_j(\lambda),
\end{eqnarray*}
where $V_j \in \{A_j$, $B_j\}$, and $\rho(\lambda)$ is the statistical
distribution of $\lambda$ with satisfying $\rho(\lambda) \ge 0$ and
$\int d\lambda \rho(\lambda) = 1$. The classical upper bound of the Bell
function $\mathcal{B}$ will be obtained by noting the following facts.
Firstly, by definition the values of $A_j$ and $B_j$ are
$\omega^{\alpha_j}$ and $\omega^{\beta_j}$ where $\alpha_j$ and
$\beta_j$ are integers. Secondly, for integer $\alpha$, we have two
identities, a)
$\sum_{n=0}^{d-1}\omega^{\alpha n} = d \delta_d(\alpha)$ where
$\delta_d(\alpha)= 1$ if $\alpha \equiv 0 \mod d$ and
$\delta_d(\alpha)=0$ otherwise, and b) $\sum_{n=1}^{d-1}\omega^{(\alpha
  + \frac{1}{2})n} + c.c. = 0$. Thirdly, the average value of a function
$f(\lambda)$ must be less than its maximum: $\int
d\lambda\rho(\lambda)f(\lambda) \le \sup_\lambda f(\lambda)$.
We then obtain
\begin{eqnarray*}
&\mathcal{B}& \le\frac{d}{4}\Big[(\delta_d(\alpha_1 + \alpha_2 +
\alpha_3) + \delta_d(\alpha_1 + \beta_2 + \beta_3 + 1) \nonumber\\ 
&+& \delta_d(\beta_1 + \alpha_2 + \beta_3 + 1) + \delta_d(\beta_1 +
\beta_2 + \alpha_3 + 1)\Big] -1. 
\end{eqnarray*}
If $d$ is even, the four arguments in the $\delta_d$ functions cannot
all be $0$ modulo $d$ because $\alpha_1 + \alpha_2 + \alpha_3$ would be
even and then the sum of the last three arguments would be odd.  Hence
\begin{equation}
 \label{eq:bfbfa}
\mathcal{B} \le \frac{3d}{4} - 1,~~~ \text{ if $d$ is even}.
\end{equation}
For \emph{odd} $d$, on the other hand, the Bell function has a larger
classical upper bound, $B\le d-1$. We will show that quantum mechanics
violates the generic Bell inequalities in even dimensions while it does
not in odd dimensions.

{\em Violation by quantum mechanics.-- } For quantum mechanical
description, we introduce an operator $\hat{V}_j$ to represent the
measurement for a variable $V_j \in \{A_j,B_j\}$ of $j$-th observer. An
{\em orthogonal} measurement of a given variable $V$ is described by a
complete set of orthonormal basis vectors $\{|\alpha \rangle_{V}\}$.
Distinguishing the measurement outcomes can be indicated by a set of
values, called eigenvalues. As the variable $V \in \{A, B\}$ takes a
value $\omega^\alpha \in S$, let the eigenvalues be the elements in $S$
so that the operator is represented by $\hat{V} = \sum_{\alpha=0}^{d-1}
\omega^\alpha |\alpha\rangle_{V V} \langle \alpha|$.  In this
representation the ``observable'' operator $\hat{V}$ is unitary
\cite{Lee04,Cerf02,Brukner02}.  Each measurement described is
nondegenerate with all distinct eigenvalues, called a maximal test
\cite{Peres98}.

The statistical average over the local hidden variable $\lambda$ in
Eq.~(\ref{eq:bellfun}) is replaced by a quantum average to derive the
quantum mechanical Bell function. For the purpose we obtain the $n$-th
order quantum correlation function
\begin{eqnarray*}
  \label{eq:nocfqm}
  E^n_{V_1 V_2 V_3} = \langle \psi | \hat{V}^n_1 \otimes \hat{V}^n_2
  \otimes \hat{V}^n_3 |
  \psi \rangle, 
\end{eqnarray*}
where $\hat{V}_j^n$ is the $n$-th power of $\hat{V}_j$ and
$|\psi\rangle$ is a quantum state of the system. It is obvious that
$|E^n| \le 1$ as the operator $\hat{V}_j$ and its powers are all
unitary. After replacing the local hidden-variable averages for $A_j$
and $B_j$ with the quantum averages for $\hat{A}_j$ and $\hat{B}_j$ in
Eq.~(\ref{eq:bellfun}), we derive the quantum mechanical Bell function
in a useful form of
\begin{eqnarray}
\label{eq:bellfun2}
\mathcal{B}_q = \frac{1}{2^2} \sum_{n=1}^{d-1} &\Big(E^n_{A_1A_2A_3} +
  \omega^n E^n_{A_1B_2B_3} \nonumber \\ 
  +& \omega^n E^n_{B_1A_2B_3} + \omega^n E^n_{B_1B_2A_3} \Big).
\end{eqnarray}
By noting $|E^n| \le 1$, the generalized triangle inequality implies
that $\mathcal{B}_q$ is bounded from above,
\begin{eqnarray}
  \label{eq:qubobf}
  |\mathcal{B}_q| \le d-1.
\end{eqnarray}
The quantum upper bound $(d-1)$ is larger than the classical upper bound
$(3d/4-1)$ for an even dimension, as shown in (\ref{eq:bfbfa}), while
they are equal for an odd dimension.  It is not clear yet whether
$\mathcal{B}_q$ actually takes the quantum upper bound as its maximum.
If so, it implies that quantum mechanics violates the generic Bell
inequalities~(\ref{eq:bfbfa}) for even $d$'s that any local realistic
theories must satisfy. We will show that this is indeed the case if the
system is prepared in a generalized GHZ state and the observable
operators $\hat{A}_j$ and $\hat{B}_j$ are given by ones which are
employed in showing a generalized GHZ nonlocality \cite{Lee04}.

A generalized GHZ state for a tripartite $d$ dimensional system is
defined as
\begin{eqnarray}
  \label{eq:ghzstate}
  |\psi \rangle =\frac{1}{\sqrt{d}} \sum_{\alpha=0}^{d-1}
|\alpha,\alpha,\alpha\rangle
\end{eqnarray}
where $\{|\alpha\rangle\}$ is an orthonormal basis set.  We consider the
two observable operators $\hat{A}$ and $\hat{B}$ that are introduced in
Ref.~\cite{Lee04}. For a given eigenvalue $\omega^\alpha$, the
eigenvector of $\hat{A}$ is given by applying quantum Fourier
transformation $\hat{F}$ on the basis vector $|\alpha\rangle$ in
Eq.~(\ref{eq:ghzstate}):
\begin{eqnarray}
  \label{eq:obp1}
  |\alpha\rangle_{A} = \hat{F}|\alpha\rangle =
   \frac{1}{\sqrt{d}} \sum_{\beta=0}^{d-1} \omega^{- \alpha
   \beta}|\beta\rangle,
\end{eqnarray}
where the subscript $A$ stands for the observable $\hat{A}$. Similarly,
the eigenvector of $\hat{B}$ is given by
\begin{eqnarray}
  \label{eq:obp2}
  |\alpha\rangle_{B} = \hat{P}_{1/2}\hat{F}|\alpha\rangle =
  \frac{1}{\sqrt{d}} \sum_{\beta=0}^{d-1}
  \omega^{-(\alpha+\frac{1}{2})\beta}|\beta\rangle
\end{eqnarray}
where $\hat{P}_\nu$ is a phase shift operator such that $\hat{P}_\nu
|\alpha\rangle = \omega^{-\nu \alpha} |\alpha\rangle$. If $d=2$, the two
observable operators $\hat{A}$ and $\hat{B}$ reduce to Pauli operators
$\hat{\sigma}_x$ and $\hat{\sigma}_y$, respectively.
It is convenient to introduce a raising operator $\hat{J}$ such that
$\hat{J} |d-1\rangle = 0$ and $\hat{J} |\alpha\rangle =
|\alpha+1\rangle$ for $0\leq\alpha < d-1$. In particular the raising
operator can be written in terms of $\hat{A}$ and $\hat{B}$ as $\hat{J}
= ( \hat{A} + \omega^{1/2} \hat{B} )/2$.  Its Hermitian conjugate
($h.c.$), $\hat{J}^\dag$, is a lowering operator. The operator
$\hat{J}^n$ ($\hat{J}^n {}^\dag$) implies raising (lowering) by $n$
levels. It is remarkable that the $n$-level raising operator can be
expressed by the $n$-th powers of $\hat{A}$ and $\hat{B}$, that is,
$\hat{J}^n = (\hat{A}^n + \omega^{n/2} \hat{B}^n )/2$.
When $\mathcal{B}_q$ is written in the form similar to
Eq.~(\ref{eq:bellfun}), the generic Bell operator is now given by
\begin{eqnarray}
  \label{eq:gbfqm}
  \hat{\mathcal{B}}_q = \sum_{n=1}^{d-1}
  \overset{3}{\underset{j=1}{\bigotimes}} 
  \hat{J}_j^n +  h.c.
\end{eqnarray}
The generalized GHZ state $|\psi\rangle$ is an eigenstate of
$\hat{\mathcal{B}}_q$ with the eigenvalue $(d-1)$, i.e., $\mathcal{B}_q
= \langle \psi | \hat{\mathcal{B}}_q |\psi\rangle = d-1$. It implies
that the quantum expectation $\mathcal{B}_q$ takes as its maximum the
quantum upper bound $(d-1)$, derived in Eq.~(\ref{eq:qubobf}), if the
tripartite system is in the generalized GHZ state $|\psi\rangle$ and the
measurements are chosen with their bases in Eqs.~(\ref{eq:obp1}) and
(\ref{eq:obp2}).  We have shown {\em the Bell theorem for a tripartite
  even dimensional system that quantum mechanics conflicts with any
  local realistic description.}

In showing the Bell theorem, the maximal tests of measurements enable
the quantum expectation $\mathcal{B}_q$ to take the quantum upper bound
in Eq.~(\ref{eq:qubobf}). As the violations are being exhibited in even
dimensions, nevertheless, one might apprehend that our Bell inequalities
would be simple extensions of or equivalent to the ones with dichotomic
observables, for instance, Mermin's inequality \cite{Mermin90}. This is
not the case: No dichotomic observables can achieve the quantum upper
bound (\ref{eq:qubobf}).
Further, if the observables are simultaneously decomposable into the
direct sum of dichotomic observables, the inequality is equivalent to a
two dimensional one \cite{Lee04,Cerf02,Cabello02}. However, our generic
Bell inequalities are genuinely $d$ dimensional in the sense that the
observables are not simultaneously decomposable into any subdimensional
observables \cite{Lee04}.

{\em Generalization to multipartite systems.--} The Bell theorem for a
tripartite even dimensional system has been proven by showing the
incompatibility of the quantum expectation with the generic Bell
inequality~(\ref{eq:bfbfa}).  Our formulation can be generalized to {\em
  arbitrary} multipartite {\em even} dimensional systems simply by
increasing the number of parties, $N \ge 3$. Then, the Bell inequality
becomes $\mathcal{B} \le d (2^{-N/2} + 2^{-1})-1$ if $N$ is even, and
$\mathcal{B} \le d (2^{-(N+1)/2} + 2^{-1})-1$ otherwise. The quantum
expectation $\mathcal{B}_q$ is independent of $N$ and it takes the
maximum of $(d-1)$, which is clearly larger than the upper bounds of the
Bell inequalities. For qubits of $d=2$, our generic Bell inequalities
reduce to Mermin's \cite{Mermin90}.

For a bipartite and/or odd dimensional system, on the other hand,
quantum mechanics does not violate the generic Bell inequalities as the
classical upper bounds are equal to the quantum.  For such a system,
however, one may consider variants from the generic Bell inequalities
and show that some of them are violated by quantum mechanics, as done
for qubits \cite{Ardehali92}.

{\em Variants from the generic Bell inequalities.--} A large set of
variants can emerge from the generic Bell operators. Such a variant Bell
operator is written in the form of
\begin{eqnarray}
  \label{eq:vgbo}
  \hat{\mathcal{B}}_\nu = \sum_{n=1}^{d-1} \omega^{\nu n}
  \overset{N}{\underset{j=1}{\bigotimes}} \hat{J}_j^n  + h.c.
\end{eqnarray}
where $\nu$ is a rational number. We obtain the variant Bell operators
from generic one by some local unitary transformation, for instance,
$\hat{\mathcal{B}}_\nu$ is obtained by applying the phase shift
operation on $\hat{J}_N$: $\hat{P}_\nu^\dag \hat{J}_N^n \hat{P}_\nu =
\omega^{\nu n}\hat{J}_N^n$, where $\hat{P}_\nu$ is defined after
Eq.~(\ref{eq:obp2}). These variants have the same quantum expectation of
$(d-1$) like the generic Bell operators that they originate from.

All variants of Bell inequalities are not violated by quantum mechanics.
For instance, if $\nu$ is an integer, the variant becomes equivalent to
the generic Bell inequality, which is not violated in a bipartite
system. For any $N$-partite $d$ dimensional system, consider the variant
Bell functions, $\mathcal{B}_{1/4}^o$ for an odd $N$ and
$\mathcal{B}_{1/4}^e$ for an even $N$, that satisfy the inequalities,
\begin{eqnarray}
  \label{eq:bifaso}
  \mathcal{B}^o_{1/4} \le && \frac{1}{2^{N-1}} \sum_{k=0}^{\frac{N-1}{2}}
      b_{N,k}-1, \\
  \label{eq:bifase}
  \mathcal{B}^e_{1/4} \le && \frac{1}{2^{N}} \left( \sum_{k=0}^{\frac{N-2}{2}}
      b_{N+1,k} + b_{N+1,\frac{N}{2}} \right)-1,
\end{eqnarray}
where $b_{n,k} = (-1)^k \binom{n}{(n-1-2k)/2} \cot(\pi(2k+1)/4d)$ with a
binomial coefficient $\binom{a}{b}$. The classical upper bounds in the
inequalities~(\ref{eq:bifaso}) and (\ref{eq:bifase}) are smaller than
the quantum expectation, $(d-1)$. Thus, {\em the Bell theorem holds for
  all multipartite arbitrary dimensional systems}. For qubits, the
variant inequalities are equivalent to those derived in
Ref.~\cite{Ardehali92}.  For a bipartite system of $N=2$, in particular,
the inequality~(\ref{eq:bifase}) becomes
\begin{eqnarray}
  \label{eq:bifq}
  \mathcal{B}^e_{1/4} \le \frac{1}{4}\left(3 \cot\frac{\pi}{4
  d}-\cot\frac{3\pi}{4 d} \right)-1. 
\end{eqnarray}
Further a two-qubit system has the classical and quantum upper bounds of
$1/\sqrt{2}$ and unity, respectively, and thus their ratio is equal to
$\sqrt{2}$ as is in the Clauser-Horne-Shimony-Holt inequality
\cite{CHSH69}.

{\em Remarks.--} In our Bell inequalities, the quantum expectation to
classical upper bound ratios (QCRs) are always larger than unity, which
is a clear proof of the Bell theorem for all the systems.  The QCRs in
our Bell inequalities decrease as the dimensionality grows, which should
be compared with the opposite behavior of the Bell inequality that
Collins {\em et al} proposed for a bipartite arbitrary dimensional
system \cite{Collins02}. Acin {\em et al.} \cite{Acin04} also suggested
an inequality to show a similar trend for a tripartite three dimensional
system.  It has been shown \cite{Son04} that Collins {\em et al.}'s
composite measurements are classically correlated. On the other hand,
those in this Letter are composed of mutually independent local
measurements. The potential inconsistency is in fact not a problem as
Bell inequalities are not uniquely defined in any way.  However, one
important issue to point out is that whereas the inequalities by the
others are maximally violated for the {\em partially} entangled state
\cite{Acin02}, our Bell inequalities show the maximum violation for the
{\em maximally} entangled state.  It is an open question whether one can
construct other variants of Bell inequalities which show the increase of
the QCR for the increase of dimensions.

In summary, we proposed the Bell inequalities for all multipartite
arbitrary dimensional systems. They were constructed based on the
generalized GHZ nonlocality for multipartite multidimensional systems
\cite{Lee04}. Quantum mechanics violates the generic Bell inequalities
for even dimensional $N$-partite systems for $N \ge 3$. The generic Bell
inequalities were shown to be genuinely multidimensional in the sense
that the observables are not simultaneously decomposable into any
subdimensional ones. It was found that the large set of variants of Bell
inequalities naturally emerge from the generic ones and a particular
variant is violated by quantum mechanics for every $N$-partite $d$
dimensional system.

\acknowledgements

We acknowledge useful discussions with {\v C}. Brukner and financial
support by the Korean Ministry of Science and Technology through Quantum
Photonic Science Research Center, the KRF (2003-070-C00024) and the UK
EPSRC.

\end{document}